# Role of defects and geometry in the strength of polycrystalline graphene

Zhigong Song, Jian Wu and Zhiping Xu[*]

Applied Mechanics Laboratory, Department of Engineering Mechanics and Center for Nano and Micro Mechanics, Tsinghua University, Beijing 100084, China

[*]Corresponding author, Email: xuzp@tsinghua.edu.cn

**Abstract**

Defects in solid commonly limit mechanical performance of the material. However, recent measurements reported that the extraordinarily high strength of graphene is almost retained with the presence of grain boundaries. We clarify in this work that lattice defects in the grain boundaries and distorted geometry thus induced define the mechanical properties characterized under specific loading conditions. Atomistic simulations and theoretical analysis show that tensile tests measure in-plane strength that is governed by defect-induced stress buildup, while nanoindentation probes local strength under the indenter tip and bears additional geometrical effects from warping. These findings elucidate the failure mechanisms of graphene under realistic loading conditions and assess the feasibility of abovementioned techniques in quantifying the strength of graphene, and suggest that mechanical properties of low-dimensional materials could be tuned by implanting defects and geometrical distortion they leads to.

Graphene features a two-dimensional honeycomb lattice with $sp^2$ carbon-carbon bonds that yields excellent mechanical properties such as an ideal tensile strength of 120 GPa [1]. This unprecedingly high value holds great promises in developing novel nanoelectromechanical devices [2, 3], and high-performance nanocomposites where mechanical resistance of single sheet could be mapped to the bulk [4]. However, as commonly concerned in non-single-crystalline materials, the limited size of synthesized single-crystalline graphene domain [5, 6] raises a critical question that how could defects in polycrystalline graphene modify its mechanical performance [7-11]? Due to technical challenge in directly measuring the mechanical strength of isolated graphene by applying uniaxial or equibiaxial tensile loads, nanoindentation tests are widely performed by several groups [1, 12-16]. One of the recent experimental measurements indicates that the grain boundaries (GBs) reduces the indentation fracture force by at most 20-40% [14]. Another report shows that this reduction depends on the lattice mismatch angles at GBs and large-angle GBs have higher strengths than their low-angle counterparts [15]. Ripping processes of graphene containing GBs under electron beam irradiation is also resolved [17]. Although the pattern of cracks has been identified in these experiments, direct observation of the fracture nucleation and propagation, and thus correlation between material strength and the detailed atomic structure of GBs has not been discussed [14, 15, 17]. More specifically, the lack of direct local stress measurement prohibits reliable characterization of the strength for polycrystalline graphene. We point out in this work that, the nanoindentation test, although has been assessed by reporting the intrinsic strength of pristine graphene and quantifying the strength reduction at GBs, probes local strength of the lattice under the indenter tip. Moreover, geometrical effects from out-of-plane deformation induced by topological defects play an important role. As a result, the fracture force in indentation cannot be directly converted to the material strength for a two-dimensional membrane with such defects subjected to in-plane loads as it usually experiences in relevant applications.

To clarify the deformation and failure mechanisms of graphene under various loading conditions, and quantify the relation between indentation fracture force and in-plane material strength, we carry out classical molecular dynamics (MD) simulations based on the adaptive intermolecular reactive empirical bond-order (AIREBO) potential [18] to explore mechanical properties of pristine and polycrystalline graphene under both nanoindentation and equibiaxial tensile loads. For polycrystalline graphene, we focus on the effect of topological defects present in GBs only. In chemical vapor deposition (CVD) graphene growth, the defects include topological pentagon-heptagon (5|7) defect pairs that can also be considered as dislocations, as well as a combination of low-mass-density defects such as octagons, vacancies, and nanovoids [15, 19, 20]. The paired 5|7 defects have been evidently observed in tilt GBs and are reported to induce polar and high-amplitude stress buildup in graphene (~56.83 GPa) [15, 20]. According to the 2D edge dislocation theory by neglecting the out-of-plane lattice distortion that releases in-plane stress, a pileup of 5|7 dislocations leads to accumulated stress buildup [7, 9, 20]. This stress buildup decays from the dislocation core as $1/r$ at a distance $r$. A finite GBs containing 5|7 pairs with length $l$ yield a tensile (compressive) stress buildup $\sigma \sim \log_{10}l$ at its end terminated with a heptagon (pentagon) [7]. Other types of defects in the GB, such as octagons, vacancies and voids, have also been identified, although less frequently, in experimental observations [21]. These defects have reduced in-plane mass densities compared to the hexagonal lattice and relatively higher formation energies compared to the topological defects. In this work, we set our focus on the topological defects with 5|7 dislocations.

Nanoindentation tests are performed using MD simulations to quantify the effects of the aforementioned defects in modifying strength of graphene. A graphene sheet is deposited to an adhesive porous substrate with the pore diameter $2R = 11$ nm (**Fig. 1**). The indenter is simulated by a spherical particle (the diameter $D = 2$ nm) placed on top of the GBs, interacting with the graphene through a harmonic spring potential with a spring stiffness $k$ of 10 eVÅ$^{-3}$. As shown in **Fig. 1a**, two types of straight GBs are firstly explored, including the armchair-oriented GB (aGB) and zigzag-oriented GB (zGB) that correspond to tilt angles of 28.7º and 21.7º [10]. Graphene with straight GBs across the membrane features is relatively flat, with the out-of-plane displacement below 0.3 nm. The results show that when the indenter is pressed downward the center of graphene membrane, the reduction of strengths, measured as the indentation force at fracture, are 51.92 and 43.95 nN for aGB and zGB, showing reduction of 14.03% and 27.22% from the value for a pristine graphene (60.39 nN). In a second set of simulations, we consider

V-shaped GBs with an angle of 2π/3 (**Fig. 1b**), the fracture force for aGB and zGB are 51.89 and 43.48 nN, with reduction of ideal strength by 14.08% and 28%. These results show consistence with recent experimentally measured reduction in the range of 20-40% [14, 15]. The fracture of polycrystalline graphene nucleates from lattice under the indenter, by breaking C-C bonds at the boundary between heptagon and hexagon. After this critical point, the crack propagates, resulting in a fracture pattern along the radial directions. The same phenomena are observed for both straight and V-shape GBs.

We further carry out nanoindentation tests by shifting the tip of indenter laterally from the center of graphene membrane by the size of indenter, i.e. 2 nm (see **Fig. 1c** and **1d** for illustration). The results show that the intrinsic strength of pristine graphene is recovered even with the same GBs across the membrane, with measured amplitudes for the fracture force as 60.11 and 59.82 nN for straight aGB and zGB, and 59.69 and 59.90 nN for V-shape aGB and zGB, respectively. In nanoindentation tests for a membrane, the in-plane tensile stress decreases inversely with distance from the indenter tip, and thus the stress under the tip at rupture describes local strength and the measured value is very sensitive to the position of indenter tip. This dependence may explain the diverse values reported in the literature [1, 12-14], in addition to the effect of wide distribution in defect types within the GBs.

Tensile stress buildup from dislocation pileups in the GBs could lead to prominent strength reduction under uniaxial or equibiaxial loads [7]. According to our discussion above, this reduction may not be detected under nanoindentation with the probe displaced away from the buildup. To verify this argument, we carry out indentation tests for graphene membranes with semi-infinite GBs and embedded GBs. The results are summarized in **Fig. 2**. We firstly explore semi-infinite GBs with different lengths, but all originating from the supported side with graphene-substrate contact and ending with a pentagon. The measured fracture force under indentation shows a peak (68.69 and 77.01 nN for aGB and zGB) when the end of GBs is located in the center of membrane where the indenter tip is placed. That is, the length of GB is $l$ = 5.57 and 5.10 nm for aGB and zGB, respectively. It should be noticed that this peak of fracture forces is 13.74% (aGB) or 27.52% (zGB) higher than that measured for pristine graphene membrane. This can be explained as a geometrical effect from the conical membrane shape formed due to the presence of pentagon, which can be considered as a positive disclination [20], as shown in **Fig. 2b** and **2c**. Now consider a conical membrane experiencing an indentation force $f$ at the bottom (**Fig. 3**), which can be separated into two parts, including a conical part free of normal load and a spherical part conformed to the spherical indenter. Arguments can then be made based on axial equilibrium in the non-contact part. At position $r$ measured laterally from the center of membrane, the meridional stress $\sigma_m(r)$ can be related to $f$ as $f = \sigma_m t 2\pi r \sin\varphi$, where $t$ is the thickness of membrane thickness and $\alpha = \pi - 2\varphi$ is the conical angle. As a result, with the same $f$ applied, the amplitude of $\sigma_m$ decreases as the conical angle decreases, or the height of the cone $d_0$ increases. By assuming linear in-plane elastic response of graphene, one can see that the effective stiffness $k = \partial f/\partial d$ increases with the cone height $d_0$ as $k = 2\cos\varphi d_0/[D\ln(2R/D\sin\varphi)]$. Under the indenter tip, $\sigma_m$ increases and circumferential stress arises due to the presence of pressure from indentation. As a result, fracture in the membrane tends to nucleate under the tip and the fracture force increases with $d$.

For very short semi-infinite GBs, the graphene lattice under the indentation tip is perfect and thus the fracture force measured is close to that of a pristine graphene. While for GBs with length close to diameter of the supporting pore, the measured strength reaches the value for straight GBs as we discussed earlier accordingly. Semi-infinite GB terminated by a heptagon is also investigated. The out-of-plane distortion does not allow the graphene membrane to be well adhered to the substrate and the structure becomes mechanically instable due to the presence of high tensile stress at its end [7]. This makes quantitative discussion on the strength of semi-infinite GBs with heptagon termination not feasible.

For the embedded GB with 5|7 pairs shown in **Fig. 2e** and **2f**, the two ends are terminated by the pentagon and heptagon. The pre-tension built up at the bond between heptagon and hexagon leads to significant reduction of the measured strength. As the length of GB $l$ increases and the leading heptagon moves away from the position of indentation, the measured strength increases to a peak with the aforementioned geometrical effect. Beyond this length, the amplitude of out-of-plane distortion is reduced and the fracture force decays to the value of a straight GB across the whole membrane. Embedded GBs with two ends terminated by pentagons or heptagons only are also investigated, but the

results show irregular dependence on the length of GB due to the magnified warping of the membrane.

To further explore the combined effect from stress buildup by 5|7 dislocations and distorted planar geometry, we construct a set of polycrystalline graphene sheet containing triple-GB junctions, as shown in **Fig. 3a**. The results from indentation tests are summarized **Fig. 3b-e**, which display distinct dependence of the stiffness and strength on the membrane morphology at rest. Specifically, we observe a positive correlation between the indentation strength, stiffness and ultimate indentation depth before fracture $d = d_0 + \Delta d$, where $d_0$ is the depth due to the presence of topological defects and $\Delta d$ is the displacement induced by the indentation force. This observation is consistent with our previous discussion of the geometrical effect on the *f-d* dependence. From the simulation results, we find that for junction with a pentagon/heptagon at the indentation position, the fracture force is enhanced/reduced. The strength and stiffness of the membrane gradually changes from the values for a single pentagon/heptagon under indentation, approaching the value for pristine graphene with a hexagon there. These results clearly indicate the combined effect of local lattice imperfections and geometry in defining the indentation strength of graphene with topological defects.

In polycrystalline graphene grown by CVD, GBs form when neighboring flakes meet up. The atomic structures of GBs, their junctions, and defects therein are defined by the formation energies, as well as local constraints such as stress, substrate adhesion, and density of carbon sources [15, 19-21]. Our MD simulation results shown above suggest that the strength measured from nanoindentation not only depends on the type of defects in the GB that modulates stress buildup topologically in the membrane, but also geometrical effect that could lead to modified (either enhanced or reduced) strength than that for pristine graphene due to the presence of out-of-plane distortion. A direct conclusion from these findings is that the fracture force, and stiffness, measured in nanoindentation of supported graphene membranes may not be able to capture the material strength that is concerned in practical applications, such as nanoeletromechanical devices or nanocomposites, where in-plane tensile loads are imposed. Additional information about the location, distribution of the defects, and the morphological warping thus induced should be taken into account for.

This conclusion can be further illustrated by considering a pristine graphene sheet consisting of a hexagonal 5|7 dislocation loop at a distance $r$ from the center of membrane [22, 23]. We find that, as shown in **Fig. 4**, the strength measured from nanoindentation and equibiaxial loading tests features contrastive $r$-dependence. In nanoindentation, the fracture nucleates from the center of membrane under the indentation tip (**Fig. 4c**) and the strength measures the critical indentation force of perfect graphene lattice with perturbation from the dislocation loop around. As the stress induced by the dislocations decays radially from the dislocation, the measured strength increases as the size of defect loop $a$ ($= r$) increases and approaches the value for pristine graphene. In contrast, in the equibiaxial tensile tests, the strength of the material is defined by the maximum tensile stress buildup in the membrane, which increases with the size of defect loop following the dislocation pileup mechanism [7]. As a result, the fracture nucleates from the defects (**Fig. 4d**) the strength measured decreases with $a$, in opposite to the indentation test.

It should be noted that the indenter used in previous discussion is frictionless. When the friction between the indenter and membrane is considered, the membrane stress in graphene under the indenter tip could be reduced by the contribution from shear stress at the interface that scales with the contact area. As a result, the lattice imperfections under the tip with weakened stress could be retained and fracture of the membrane nucleate at the boundary of contact instead. Also the spherical indenter with limited size here may differ from the realistic situation in nanoindentation tests and the length scale of both locality in strength probe and geometrical effect should be adjusted when referring to experimentally measured data quantitatively.

The dual roles of GBs in defining the strength of pristine or defected graphene have thus been assessed. The local stress buildup reduces its strength, and local buckles due to topological defects modulate it under nanoindentation, in contrast to the situation where in-plane loads are imposed. These findings elucidate the possible mechanisms featured by reported simulation and experimental results, and also raise the question that how the fracture force measured by nanoindentation could be mapped to strength of the material under tensile loading conditions that are more common in practical applications. Other techniques, e.g. the blister test [24, 25], may serve as a better choice by avoiding stress localization in the membrane under loading although the contact between the membrane and the supports should be carefully designed to avoid fracture nucleation therein. Moreover, the notable defect and geometry effects uncovered here imply that material properties of low-dimensional materials such as graphene could be tuned to a large extent by simply implanting topological defects that result in stress localization and geometrical distortion.

**Acknowledgment**

This work was supported by the National Natural Science Foundation of China through Grant 11222217, 11002079, and Tsinghua University Initiative Scientific Research Program 2011Z02174. The computation was performed on the Explorer 100 cluster system of Tsinghua National Laboratory for Information Science and Technology.

**Figures and Captions**

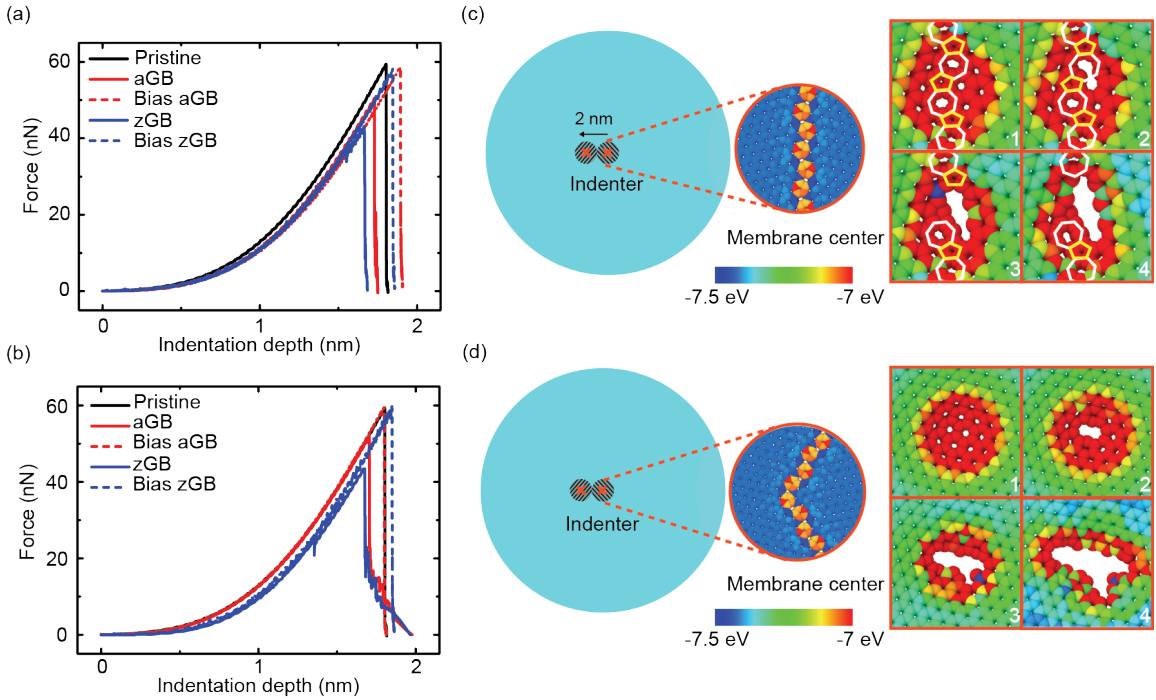

**Fig. 1.** Nanoindentation tests using MD simulations. (a) and (b) show the indentation force-depth relations for graphene membranes with straight and V-shape GBs. The atomic structures of GBs in the center of supported polycrystalline membrane, as well as the fracture patterns, are plotted in panels (c) and (d), respectively. Nanoindentation is carried out in the center or membrane or with a bias of 2 nm. Numbers in panels (c) and (d) indicate the sequence during the fracture process. Colors in panel (c) and (d) depict the potential energy per atom, which will be used in following figures as well.

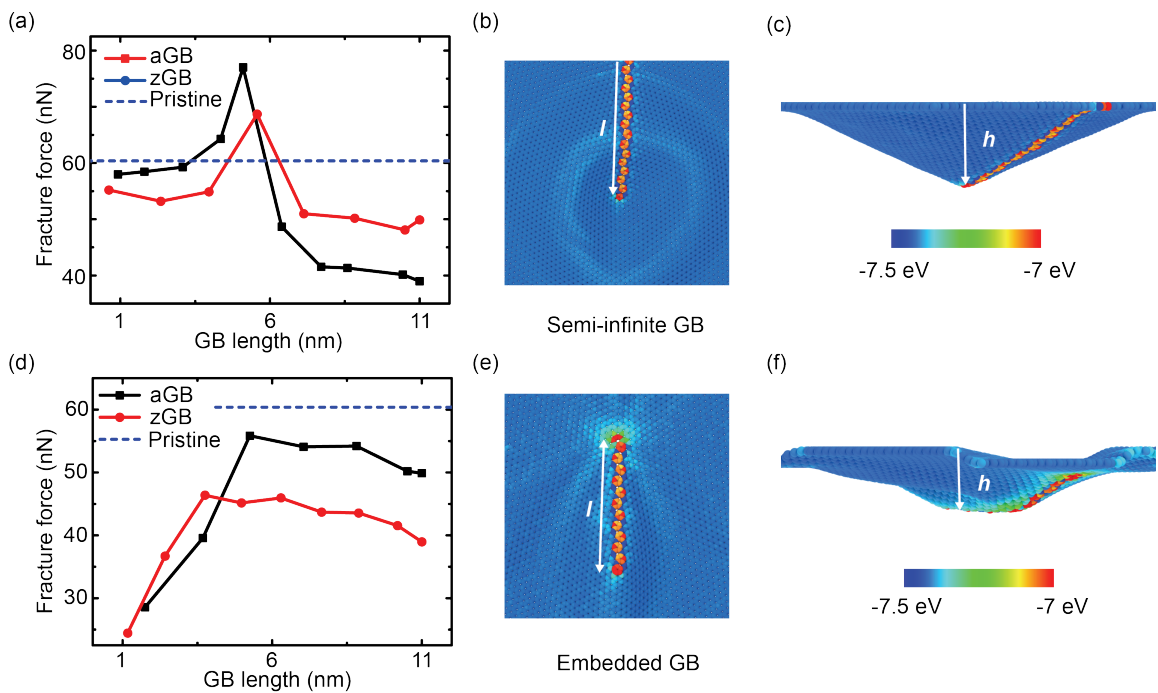

**Fig. 2.** Indentation tests on (a,b,c) semi-infinite and (d,e,f) embedded GBs. The measured fracture force shows significant length dependence, which arises from a combined effect from the tensile stress buildup in the heptagons and distorted membrane geometry (c,f).

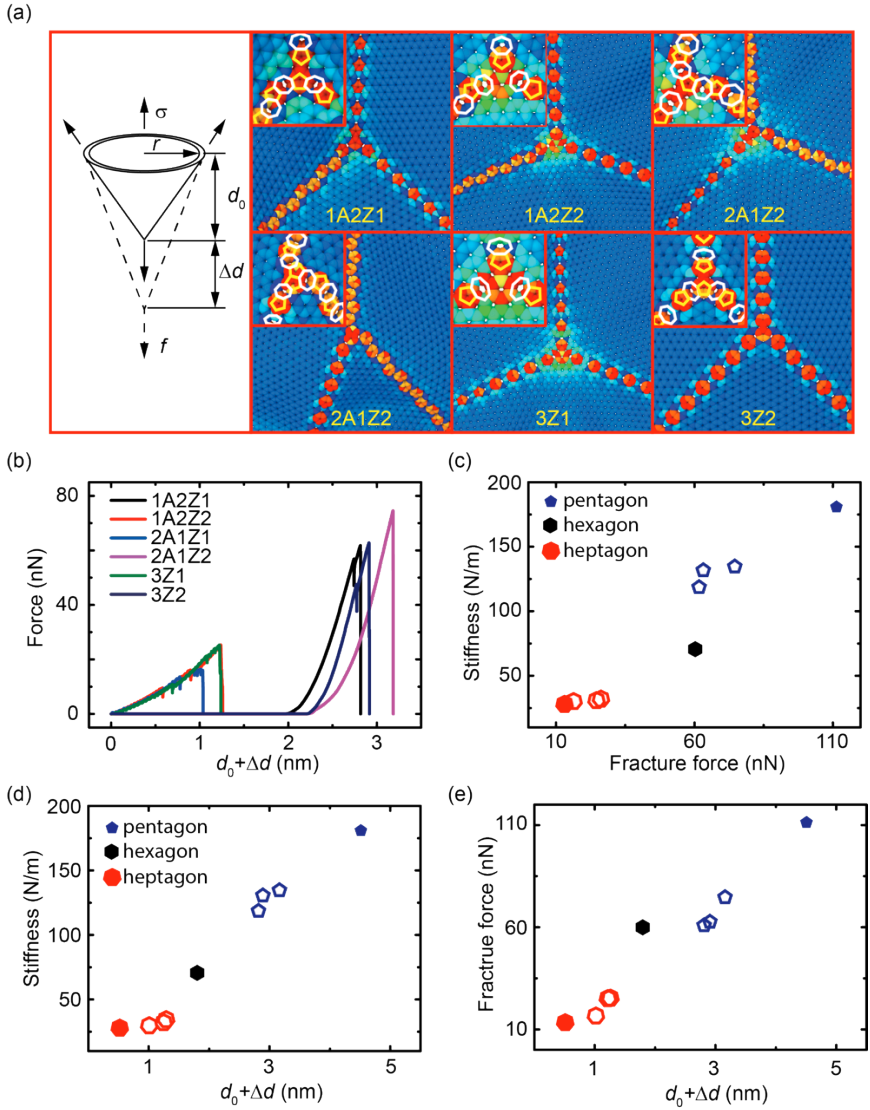

**Fig. 3.** Nanoindentation tests on triple-GB junctions. (a) The atomic structures show distinct out-of-plane buckling morphology. (b) and (c), the measured indentation strength shows significant dependence of the membrane morphology, as indicated by the correlation between fracture force, stiffness and the vertical position of indenter $d = d_0 + \Delta d$ (d,e).

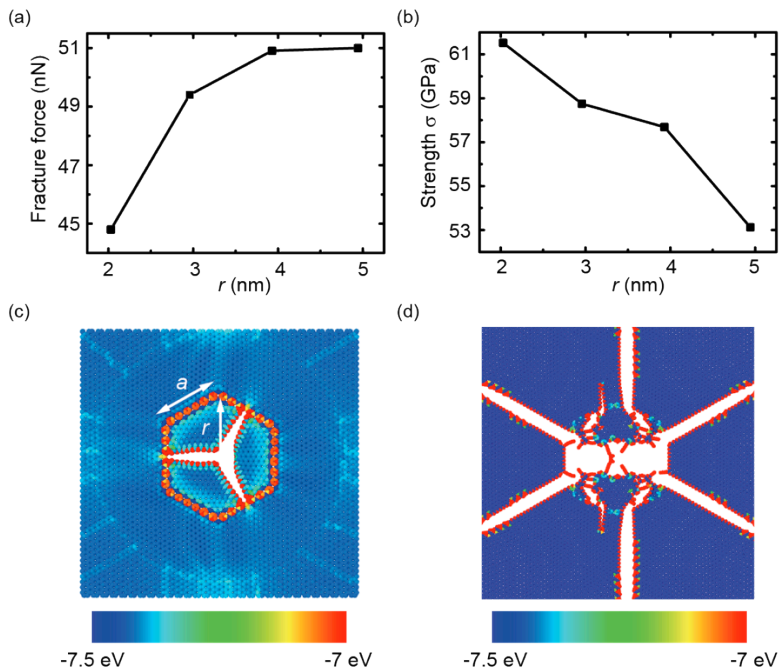

**Fig. 4.** Fracture force and tensile strength measured for a graphene sheet containing a 5|7 dislocation loop, as measured from (a) nanoindentation and (b) equibiaxial tensile tests, respectively. Panels (c) and (d) show the fracture patterns, which nucleate from the center of membrane and dislocation loop, respectively.